\documentclass[12pt]{article}
\usepackage{amsmath}
\usepackage{amssymb}
\usepackage{amsfonts}
\usepackage{graphics}

\title{Novel exactly solvable Schr\"odinger equations with a position-dependent mass in multidimensional spaces 
obtained from duality}

\author{C.\ QUESNE\thanks{E-mail address: cquesne@ulb.ac.be} \\
{\small\sl Physique Nucl\'eaire Th\'eorique et Physique Math\'ematique, 
Universit\'e Libre de Bruxelles,} \\ 
{\small \sl Campus de la Plaine CP229, Boulevard~du Triomphe, B-1050
Brussels, Belgium}}
\date{ }
\begin{document}
\baselineskip=22pt plus 1pt minus 1pt
\maketitle

\begin{abstract}
A novel exactly solvable Schr\"odinger equation with a position-dependent mass (PDM) describing a Coulomb problem in $D$ dimensions is obtained by extending the known duality relating the quantum $d$-dimensional oscillator and $D$-dimensional Coulomb problems in Euclidean spaces for $D = (d+2)/2$. As an intermediate step, a mapping between a quantum $d$-dimensional nonlinear oscillator of Mathews-Lakshmanan type (or oscillator in a space of constant curvature) and a quantum $D$-dimensional Coulomb-like problem in a space of nonconstant curvature is derived. It is finally reinterpreted in a PDM background. 
\end{abstract}

\vspace{0.5cm}


\noindent
{\sl Short title}: Exactly solvable Schr\"odinger equations with a position-dependent mas
 
\newpage
%
%
\section{Introduction}

In recent years, Schr\"odinger equations with a position-dependent mass (PDM) have attracted a lot of attention because PDM is of utmost relevance in a wide variety of physical situations, such as in energy density many-body problems, in electronic properties of semiconductors and quantum dots, in quantum liquids, ${}^3$He clusters, and metal clusters (see, {\it e.g.}, \cite{bagchi05, cq06, mustafa} for a list of references). As in constant-mass cases, exact solutions play an important role because they may provide both a conceptual understanding of physical phenomena and a testing ground for some approximation schemes. As a consequence,  several methods have been tried to derive them.\par
%
%
It is well known that Schr\"odinger equations in a PDM background may be alternatively interpreted as Schr\"odinger equations in curved spaces. On the other hand, a nonlinear oscillator in a space of constant curvature
\cite{carinena07b, carinena07c, carinena12, cq15}, which has arisen as a quantization \cite{carinena04a, carinena07a} of the celebrated Mathews and  Lakshmanan one-dimensional classical nonlinear oscillator \cite{mathews}, has attracted much interest. The aim of this communication is to show that it may serve as a starting point for building new exactly solvable PDM Schr\"odinger equations.\par
%
%
A clue to this construction is the relationship between the $d$-dimensional harmonic oscillator problem and the $D$-dimensional Coulomb one in Euclidean spaces for specific values of the pair $(d,D)$, resulting from coupling-constant metamorphosis \cite{hietarinta}, St\"ackel transform \cite{boyer}, regularization of the Coulomb problem \cite{levi, kustaan, hurwitz}, supersymmetry \cite{kostelecky}, or duality tranformation \cite{mittag}. Since the latter is a very simple and powerful method for generating new exactly solvable potentials (see, {\it e.g.}, \cite{bagchi04}), we choose to apply it in the present case. We plan to extend it to curved spaces, then to reinterprete the relation so obtained in a PDM background.\par
%
%
\section{Going from the oscillator to the Coulomb problem in Euclidean spaces}

As well known , the radial equation for the quantum $d$-dimensional oscillator in an Euclidean space reads
\begin{equation}
  \left(\frac{d^2}{dr^2} + \frac{d-1}{r} \frac{d}{dr} - \frac{l(l+d-2)}{r^2} - \omega^2 r^2 + 2E\right) R(r) = 0,
  \qquad 0 < r < +\infty,
\end{equation}
in units wherein $\hbar = m = 1$. Here $\omega$ and $l$ denote the oscillator frequency and the angular momentum quantum number, respectively. It has an infinite number of bound-state solutions, which, up to some normalization factor, can be written as
\begin{equation}
  R_{n_r,l}(r) \propto r^l e^{-\frac{1}{2}\omega r^2} L_{n_r}^{\left(l + \frac{d-2}{2}\right)}(\omega r^2),
  \qquad n_r=0, 1, 2, \ldots,  \label{eq:O-wf}
\end{equation}
in terms of Laguerre polynomials, and correspond to bound-state energies
\begin{equation}
  E_n = \omega \left(n + \frac{d}{2}\right), \qquad n = 2n_r + l.  \label{eq:O-E}
\end{equation}
The set of functions $R_{n_r,l}(r)$, corresponding to a given $l$ value, is orthogonal on the half-line $(0, +\infty)$ with respect to the measure $d\mu = r^{d-1} dr$.\par
%
%
On setting $r = \sqrt{R}$ and $R_{n_r,l}(r(R)) = S_{n_r,l}(R)$, we arrive at the differential equation
\begin{equation}
  \left(\frac{d^2}{dR^2} + \frac{d}{2R} \frac{d}{dR} - \frac{l(l+d-2)}{4R^2} + \frac{E}{2R} - \frac{1}{4}
  \omega^2\right) S(R) = 0,
\end{equation}
which can be interpreted as the radial equation for the quantum $D$-dimensional Coulomb problem,
\begin{equation}
  \left(\frac{d^2}{dR^2} + \frac{D-1}{R} \frac{d}{dR} - \frac{L(L+D-2)}{R^2} + \frac{Q}{R} + 2 {\cal E}\right)
  S(R) = 0,
\end{equation}
provided we set
\begin{equation}
  D = \tfrac{1}{2}(d+2), \qquad L = \tfrac{1}{2}l, \qquad Q = \tfrac{1}{2}E, \qquad {\cal E} = - 
  \tfrac{1}{8}\omega^2. \label{eq:C-O}
\end{equation}
This shows that the roles of the coupling constant and the energy eigenvalue have been exchanged. On combining eqs.~(\ref{eq:O-E}) and (\ref{eq:C-O}), we can rewrite $Q$ as
\begin{equation}
  Q = \omega \left(\nu + \frac{D-1}{2}\right), \qquad \text{where $\nu = n_r + L$},
\end{equation}
yielding $\omega = Q/\left(\nu + \frac{D-1}{2}\right)$. Inserting this expression in that of the Coulomb bound-state energies given in (\ref{eq:C-O}), we get
\begin{equation}
  {\cal E}_{\nu} = - \frac{Q^2}{2(2\nu+D-1)^2},  \label{eq:C-E}
\end{equation}
with corresponding wavefunctions obtained from (\ref{eq:O-wf}) in the form
\begin{equation}
  S_{n_r,L}(R) \propto R^L e^{- \sqrt{2|{\cal E}_{\nu}|} R} L_{n_r}^{(2L+D-2)}\left(2\sqrt{2|{\cal E}_{\nu}|} R
  \right).  \label{eq:C-wf}
\end{equation}
The set of functions $S_{n_r,L}(R)$, corresponding to a given $L$ value, is orthogonal on the half-line $(0,+\infty)$ with respect to the measure $d\mu = R^{D-1} dR$.\par
%
%
Equations (\ref{eq:C-E}) and (\ref{eq:C-wf}) coincide with well-known results for the Coulomb problem in a $D$-dimensional space. It is worth stressing that the bound-state spectrum of the $D$-dimensional Coulomb problem is related to half the spectrum of the $d$-dimensional oscillator (namely that of even angular momentum states) for some even integer $d$, defined in (\ref{eq:C-O}).\par
%
%
\section{Going from a nonlinear oscillator to a Coulomb-like problem}

The one-dimensional classical nonlinear oscillator, first considered by Mathews and Lakshmanan \cite{mathews}, can be described in terms of a Hamiltonian
\begin{equation}
  H = \frac{1}{2} (1+\lambda x^2) p^2 + \frac{\alpha^2 x^2}{2(1+\lambda x^2)},  \label{eq:H-1-class}
\end{equation}
where $\alpha$ plays the role of $\omega$ in the standard oscillator and the nonlinearity parameter $\lambda \ne 0$ enters both the potential energy term and the kinetic energy one. According to whether $\lambda > 0$ or $\lambda < 0$, the range of the coordinate $x$ is $(-\infty, +\infty)$ or $(-1/\sqrt{|\lambda|}, +1/\sqrt{|\lambda|})$.\par
%
%
The quantum version of $H$ has been obtained \cite{carinena04a, carinena07a} by replacing $\sqrt{1+\lambda x^2}\, p$ by the operator $- {\rm i} \sqrt{1+\lambda x^2}\, d/dx$, yielding
\begin{equation}
  \hat{H} = - \frac{1}{2} \left[(1+\lambda x^2) \frac{d^2}{dx^2} + \lambda x \frac{d}{dx}\right] +
  \frac{\alpha^2 x^2}{2(1+\lambda x^2)},  \label{eq:H-1-quant}
\end{equation}
which is formally self-adjoint with respect to the measure $d\mu = (1+\lambda x^2)^{-1/2} dx$. Such a Hamiltonian is exactly solvable for a $\lambda$-dependent potential parameter $\alpha^2 = \beta(\beta+\lambda)$. From now on, we will assume that $\alpha^2$ is defined in this way.\par
%
%
A $d$-dimensional generalization of the classical Hamiltonian (\ref{eq:H-1-class}) has been proposed \cite{carinena04b} in such a way that the resulting Hamiltonian
\begin{equation}
\begin{split}
  H &= \frac{1}{2} \biggl[\sum_i p_i^2 + \lambda \biggl(\sum_i x_i p_i\biggr)^2\biggr] 
       + \frac{\beta(\beta+\lambda) r^2}{2(1+\lambda r^2)} \\ 
  &= \frac{1}{2} \biggl[(1+\lambda r^2) \sum_i p_i^2 - \lambda \sum_{i<j} J_{ij}^2\biggr] 
       + \frac{\beta(\beta+\lambda) r^2}{2(1+\lambda r^2)}
\end{split}  \label{eq:H-d-class}
\end{equation}
keeps the maximal superintegrability property of the standard $d$-dimensional oscillator. In (\ref{eq:H-d-class}), all summations run over $i,j=1$, 2, \ldots, $d$, $J_{ij} \equiv x_i p_j - x_j p_i$ denotes an angular momentum component, and $r^2 \equiv \sum_i x_i^2$ with $r$ running on $(0, +\infty)$ or $(0, 1/\sqrt{|\lambda|})$ according to whether $\lambda>0$ or $\lambda<0$. Furthermore, $H$ may be interpreted as describing a harmonic oscillator in a space of constant curvature $\kappa = - \lambda$.\par
%
%
The quantization of (\ref{eq:H-d-class}) in two \cite{carinena07b, carinena07c} and three \cite{carinena12} dimensions has been studied, but it can be easily extended to $d$ dimensions. On replacing $\sqrt{1+\lambda r^2}\, p_i$ and $J_{ij}$ by the operators $- {\rm i} \sqrt{1+\lambda r^2}\, \partial/\partial x_i$ and $\hat{J}_{ij} = - {\rm i} (x_i \partial/\partial x_j - x_j \partial/\partial x_i)$, respectively, we arrive at
\begin{equation}
\begin{split}
  \hat{H} &= - \frac{1}{2} \biggl[(1+\lambda r^2) \Delta + \lambda r \frac{\partial}{\partial r} + \lambda
       \hat{J}^2\biggr] + \frac{\beta(\beta+\lambda) r^2}{2(1+\lambda r^2)} \\ 
  &= - \frac{1}{2} \biggl[(1+\lambda r^2) \frac{\partial^2}{\partial r^2} + (d-1 + d\lambda r^2) \frac{1}{r}
       \frac{\partial}{\partial r} - \frac{\hat{J}^2}{r^2}\biggr] 
       + \frac{\beta(\beta+\lambda) r^2}{2(1+\lambda r^2)},
\end{split}  
\end{equation}
with $\hat{J}^2 \equiv \sum_{i<j} \hat{J}_{ij}^2$ and $\Delta$ denoting the Laplacian in a $d$-dimensional Euclidean space.\par
%
%
The corresponding Schr\"odinger equation is separable in hyperspherical coordinates and gives rise to the radial equation
\begin{equation}
  \left((1+\lambda r^2) \frac{d^2}{dr^2} + (d-1 + d\lambda r^2) \frac{1}{r} \frac{d}{dr} - \frac{l(l+d-2)}{r^2}
  - \frac{\beta(\beta+\lambda) r^2}{1+\lambda r^2} + 2E\right) R(r) = 0,  \label{eq:NLO-S}
\end{equation}
where $\hat{J}^2$ has been replaced by its eigenvalues $l(l+d-2)$, $l=0$, 1, 2, \ldots. The differential operator in (\ref{eq:NLO-S}) is formally self-adjoint with respect to the measure $d\mu = (1+\lambda r^2)^{-1/2} r^{d-1} dr$. For $d=2$, eq.~(\ref{eq:NLO-S}) reduces to eq.~(29) of \cite{cq15}. Its solutions can be easily obtained by extending the $d=2$ approach to general $d$ values and are given by
\begin{equation}
  R_{n_r,l}(r) \propto r^l (1+\lambda r^2)^{-\beta/(2\lambda)} P^{\left(l + \frac{d-2}{2}, 
  -\frac{\beta}{\lambda} - \frac{1}{2}\right)}_{n_r}(1+2\lambda r^2), \qquad n_r=0, 1, 2, \ldots,
  \label{eq:NLO-wf}
\end{equation}
in terms of Jacobi polynomials, with corresponding energy eigenvalues
\begin{equation}
  E_n = \beta\left(n + \frac{d}{2}\right) - \frac{\lambda}{2} n(n+d-1), \qquad n = 2n_r + l.  \label{eq:NLO-E}
\end{equation}
The range of $n$ values in (\ref{eq:NLO-E}) is determined from the normalizability of the radial wavefunctions on the interval $(0, +\infty)$ for $\lambda>0$ or $(0, 1/\sqrt{|\lambda|})$ for $\lambda<0$ with respect to the measure $d\mu = (1+\lambda r^2)^{-1/2} r^{d-1} dr$. It is given by
\begin{equation}
  n = \begin{cases}
     0, 1, 2, \ldots & \text{if $\lambda<0$}, \\
     0, 1, 2, \ldots, n_{\rm max}, \quad \frac{\beta}{\lambda} - \frac{d+1}{2} \le n_{\rm max} < \frac{\beta}
         {\lambda} - \frac{d-1}{2} & \text{if $\lambda>0$}.
  \end{cases}
\end{equation}
It is worth observing that in the limit where $\beta/|\lambda|$ goes to infinity, the wavefunctions (\ref{eq:NLO-wf}) go over to (\ref{eq:O-wf}) (with $\omega$ replaced by $\beta$), due to a limit relation between Jacobi and Laguerre polynomials \cite{abramowitz}.\par
%
%
Let us now perform the same transformation $r = \sqrt{R}$ and $R_{n_r,l}(r(R)) = S_{n_r,l}(R)$ as in Sect.~2. This yields the differential equation
\begin{align}
  & \biggl((1+\lambda R)^2 \frac{d^2}{dR^2} + \frac{1}{2R} (1+\lambda R) [d + (d+1)\lambda R] \frac{d}{dR} 
       - \frac{l(l +d-2)}{4R^2}  \nonumber \\
  & \quad {}+ \frac{1}{4R} [2E - \lambda l(l+d-2)]- \frac{1}{4} \beta(\beta+\lambda) + \frac{1}{2} \lambda E
       \biggr) S(R) = 0,
\end{align}
which can be rewritten as
\begin{align}
  & \biggl[(1+\lambda R)^2 \frac{d^2}{dR^2} + \frac{D-1}{R} (1+\lambda R) \biggl(1 + \frac{2D-1}
       {2D-2}\lambda R\biggr) \frac{d}{dR} - \frac{L(L+D-2)}{R^2} \nonumber \\
  &\quad {}+ \frac{Q}{R} + 2{\cal E}\biggr] S(R) = 0,  \label{eq:NLC-S}
\end{align}
provided we set
\begin{equation}
  D = \tfrac{1}{2}(d+2), \quad L = \tfrac{1}{2}l, \quad Q = \tfrac{1}{2}[E - 2\lambda L(L+D-2)], \quad
  {\cal E} = - \tfrac{1}{8}\beta(\beta+\lambda) + \tfrac{1}{4}\lambda E.  \label{eq:NLC-NLO}
\end{equation}
It is straightforward to show that the differential operator in eq.~(\ref{eq:NLC-S}) is formally self-adjoint on the interval $(0, +\infty)$ for $\lambda>0$ or $(0, 1/|\lambda|)$ for $\lambda<0$ with respect to the measure $d\mu = (1+\lambda R)^{-3/2} R^{D-1} dR$, corresponding to a space of nonconstant curvature. In such a space, the potential $-Q/R$ may not be interpreted as a Coulomb potential, since the latter, obtained as a solution of Laplace equation, assumes a more complicated form. We will therefore refer to it in this section as a Coulomb-like potential.\par
%
%
As revealed by eq.~(\ref{eq:NLC-NLO}), the exchange of the roles of the coupling constant and the energy eigenvalue also looks less strict since $L(L+D-2)$ and $E$ make their appearance in $Q$ and $\cal E$, respectively. We can, however, proceed as in Sec.~2 and combine eqs.~(\ref{eq:NLO-E}) and (\ref{eq:NLC-NLO}) to write $Q$ as
\begin{equation}
  Q = \beta \left(\nu + \frac{D-1}{2}\right) - \lambda \left[\nu \left(\nu + D - \frac{3}{2}\right) + L(L+D-2)
  \right], \qquad \nu = n_r + L,
\end{equation}
yielding $\beta = \left(\nu + \frac{D-1}{2}\right)^{-1} \left\{Q + \lambda \left[\nu \left(\nu + D - \frac{3}{2}\right) + L(L+D-2)\right]\right\}$. Inserting this expression in that of $\cal E$ given in (\ref{eq:NLC-NLO}), we obtain after a straightforward calculation
\begin{align}
  {\cal E}_{n_r,L} &= - \frac{1}{2(2\nu+D-1)^2} \left\{Q + \lambda\left[-\nu\left(\nu+\frac{1}{2}\right) +
       L(L+D-2)\right]\right\} \nonumber \\
  & \quad \times \left\{Q + \lambda\left[-(\nu+D-1)\left(\nu+D-\frac{3}{2}\right) + L(L+D-2)\right]\right\}.
       \label{eq:NLC-E}
\end{align}  
It is worth observing here that, in contrast with the conservation of accidental degeneracies that occurs when going from the standard oscillator to the nonlinear one (see eqs.~(\ref{eq:O-E}) and (\ref{eq:NLO-E})) and which is related to the maximal superintegrability property conservation, nothing similar happens in the Coulomb-like case since eq.~(\ref{eq:C-E}) is replaced by (\ref{eq:NLC-E}). As a matter of fact, this is an important aspect of our method. Using a more complicated $\lambda$-dependent transformation might have maintained the maximal superintegrability property and, consequently, the accidental degeneracies, but it would not have given rise to any new result when going to a PDM picture.\par
%
%
{}From the radial wavefunctions (\ref{eq:NLO-wf}), we also get
\begin{equation}
  S_{n_r,L}(R) \propto R^L (1+\lambda R)^{\tau} P_{n_r}^{(\rho, \sigma)}(1+2\lambda R),
\end{equation}
where
\begin{equation}
\begin{split}
  \rho &= 2L+D-2, \\
  \sigma &= - \frac{1}{\lambda\left(\nu+\frac{D-1}{2}\right)} \left\{Q + \lambda\left[\nu^2 + (D-1)\nu
      +\frac{1}{4}(D-1) + L(L+D-2)\right]\right\}, \\ 
  \tau &= - \frac{1}{\lambda(2\nu+D-1)} \left\{Q + \lambda\left[\nu\left(\nu+D-\frac{3}{2}\right)
      + L(L+D-2)\right]\right\}.
\end{split} \label{eq:parameters}
\end{equation}
Bound-state wavefunctions, i.e., functions $S_{n_r,L}(R)$ normalizable with respect to the measure $d\mu = (1+\lambda R)^{-3/2} R^{D-1} dR$, correspond to sets of quantum numbers $(n_r, L)$ satisfying the inequalites
\begin{equation}
\begin{split}
  & n_r^2 + (2L+D-1) n_r + 2L^2 + (2D-3)L + \frac{1}{4}(D-1) < \frac{Q}{|\lambda|} \qquad \text{if 
       $\lambda<0$}, \\
  & n_r^2 + (2L+D-1) n_r + L + \frac{1}{4}(D-1)(2D-3) < \frac{Q}{\lambda} \qquad \text{if $\lambda>0$}.
\end{split}
\end{equation}
In both cases, there is only a finite number of sets fulfilling these conditions and there is at least one $(n_r=0, L=0)$ provided $Q > \frac{1}{4}(D-1)|\lambda|$ if $\lambda<0$ or $Q > \frac{1}{4}(D-1)(2D-3)\lambda$ if $\lambda>0$.\par
%
%
\section{Reinterpretation in a PDM background}

As well known, one of the main difficulties of PDM problems comes from the non-commutativity of the momentum and mass operators, which can be coped with by using the von Roos approach \cite{vonroos}, wherein the kinetic energy operator is written as
\begin{equation}
  - \frac{1}{4} \sum_i \left[m^{\xi}(\textbf x) \frac{\partial}{\partial x_i} m^{\eta}(\textbf x)
  \frac{\partial}{\partial x_i} m^{\zeta}(\textbf x) + m^{\zeta}(\textbf x) \frac{\partial}{\partial x_i}
   m^{\eta}(\textbf x) \frac{\partial}{\partial x_i} m^{\xi}(\textbf x)\right]
\end{equation}
in terms of some ambiguity parameters $\xi$, $\eta$, $\zeta$, constrained by the condition $\xi + \eta + \zeta = -1$. This form contains as special cases all the proposals that have been made in the literature and whose usefulness may depend on the physical problem in hand. To be more specific, we are going to consider here two special choices, namely the BenDaniel and Duke (BD) one ($\xi = \zeta = 0$, $\eta = -1$) \cite{bendaniel} and the Mustafa and Mazharimousavi (MM) one ($\xi = \zeta = -1/4$, $\eta = -1/2$) \cite{mustafa}.\par
%
%
Considering first the $d$-dimensional nonlinear oscillator described by radial equation (\ref{eq:NLO-S}), it is straightforward to show that it can be reinterpreted as a $d$-dimensional nonlinear oscillator with a PDM $m(r) = (1+\lambda r^2)^{-1}$, the BD and MM radial Schr\"odinger equations being
\begin{align}
  & \left(-\frac{d}{dr} \frac{1}{m(r)} \frac{d}{dr} + V_1(r) - 2E_1\right) \tilde{R}_{n_r,l}(r) = 0, \nonumber \\
  & V_1(r) = \frac{\left(l+\frac{d-1}{2}\right) \left(l +\frac{d-3}{2}\right)}{r^2} + \frac{\beta(\beta+\lambda)
      r^2 - \frac{1}{4}\lambda}{1+\lambda r^2}, \qquad 2E_1 = 2E_n - \frac{1}{4}d(d-2)\lambda,
\end{align}
and
\begin{align}
  & \left(-m^{-1/4}(r)\frac{d}{dr} m^{-1/2}(r) \frac{d}{dr} m^{-1/4}(r) + V_2(r) - 2E_2\right) 
      \tilde{R}_{n_r,l}(r) = 0, \nonumber \\
  & V_2(r) = \frac{\left(l+\frac{d-1}{2}\right) \left(l +\frac{d-3}{2}\right)}{r^2} + \frac{\left(\beta+
      \frac{\lambda}{2}\right)^2 r^2 + \frac{1}{4}\lambda}{1+\lambda r^2}, \qquad 2E_2 = 2E_1,
\end{align}
with $\tilde{R}_{n_r,l}(r) = r^{(d-1)/2} (1+\lambda r^2)^{-1/4} R_{n_r,l}(r)$ in both cases.\par
%
%
Similarly, the $D$-dimensional Coulomb-like problem, characterized by radial equation (\ref{eq:NLC-S}), becomes a $D$-dimensional (true) Coulomb problem with a PDM $M(R) = (1+\lambda R)^{-2}$. The BD and MM radial Schr\"odinger equations now read
\begin{equation}
  \left(- \frac{d}{dR} \frac{1}{M(R)} \frac{d}{dR} + U(R) - 2 {\cal E}_1\right) \tilde{S}_{n_r,L}(R) = 0,
  \label{eq:PDM-C1}
\end{equation}
and
\begin{equation}
  \left(- M^{-1/4}(R) \frac{d}{dR} M^{-1/2}(R) \frac{d}{dR} M^{-1/4}(R) + U(R) - 2 {\cal E}_2\right) 
  \tilde{S}_{n_r,L}(R) = 0,  \label{eq:PDM-C2}
\end{equation}
with the same potential
\begin{equation}
  U(R) = \frac{\left(L+\frac{D-1}{2}\right)\left(L+\frac{D-3}{2}\right)}{R^2} - \frac{Q - \frac{1}{4}(D-1)(2D-5)
  \lambda}{R},
\end{equation}
and $\tilde{S}_{n_r,L}(R) = R^{(D-1)/2} (1+\lambda R)^{-3/4} S_{n_r,L}(R)$, but different energy eigenvalues
\begin{equation}
  2{\cal E}_1 = 2{\cal E}_{n_r,L} - \tfrac{1}{16}(2D-1)(2D-5)\lambda^2, \qquad 2{\cal E}_2 = 2{\cal E}_{n_r,L} 
  - \tfrac{1}{16}(2D-3)^2\lambda^2.
\end{equation}
\par
%
%
As a final point, it is worth stressing that the PDM reinterpretation of Eqs.~(\ref{eq:NLO-S}) and (\ref{eq:NLC-S}), which were associated with some complicated measures, has converted them in some equations in spaces with a simple measure $dr$ or $dR$, which are directly applicable to those physical problems, mentioned in Sec.~1, wherein the PDM approach is relevant.\par
%
%
\section{Conclusion}

In this Letter, we have proved that the known duality relating the quantum $d$-dimensional oscillator problem to the quantum $D$-dimensional Coulomb one in Euclidean spaces, whenever $D = (d+2)/2$, can be easily extended to a quantum $d$-dimensional nonlinear oscillator of Mathews-Lakshmanan type (or harmonic oscillator in a constant curvature space). The result of the mapping is a quantum $D$-dimensional Coulomb-like problem in a space of nonconstant curvature. Going to an equivalent PDM description with respective masses $m(r) = (1+\lambda r^2)^{-1}$ and $M(R) = (1+\lambda R)^{-2}$ leads to a duality between a nonlinear oscillator and a (true) Coulomb problem. If the existence of the former has been (at least implicitly) signalled in previous papers on low-dimensional problems \cite{carinena07b, carinena07c, carinena04a, carinena07a, carinena04b}, that of the latter is a novel by-product of the simple transformation known in Euclidean spaces, which may lead to significant applications in those physical problems where the PDM concept appears.\par
%
%
Another usefulness of the new exactly solvable model in multidimensional spaces, proposed in the present paper, may be in the context of the $1/N$ expansion approximation method \cite{witten, mlodinow}.\par
%
%
Considering other duality transformations in curved spaces or in a PDM background than that connecting the oscillator-Coulomb pair would be a very interesting topic for future investigation.\par
%
%

\end{document}